\documentclass[conference]{IEEEtran}
\usepackage{cite}
\usepackage{amsmath,amssymb,amsfonts}
\usepackage[ruled,vlined]{algorithm2e}
\usepackage{algorithmic}
\usepackage{graphicx}
\usepackage{textcomp}
\usepackage{xcolor}
\usepackage[euler]{textgreek}
\usepackage{color}
\usepackage{url}
\usepackage{todonotes}
\usepackage{multirow}
\usepackage{tabularx}
\usepackage{subfig}
\usepackage{enumitem}
\usepackage{authblk}

\def\BibTeX{{\rm B\kern-.05em{\sc i\kern-.025em b}\kern-.08em
    T\kern-.1667em\lower.7ex\hbox{E}\kern-.125emX}}
    
\begin{document}
    
\title{Priority-based Fair Scheduling in Edge Computing}

\author{Arkadiusz Madej\textsuperscript{1}, Nan Wang\textsuperscript{2}, Nikolaos Athanasopoulos\textsuperscript{1}, Rajiv Ranjan\textsuperscript{3}, and Blesson Varghese\textsuperscript{1}}
\affil{
\textsuperscript{1}\textit{Queen's University Belfast, UK}; \textsuperscript{2}\textit{Durham University, UK}; \textsuperscript{3}\textit{Newcastle University, UK}\\
E-mail: amadej01@qub.ac.uk; nan.wang@durham.ac.uk; n.athanasopoulos@qub.ac.uk;\\raj.ranjan@ncl.ac.uk; b.varghese@qub.ac.uk (Corresponding E-mail)
}

\maketitle

\thispagestyle{plain}
\pagestyle{plain}

\begin{abstract}
Scheduling is important in Edge computing. In contrast to the Cloud, Edge resources are hardware limited and cannot support workload-driven infrastructure scaling. Hence, resource allocation and scheduling for the Edge requires a fresh perspective. Existing Edge scheduling research assumes availability of all needed resources whenever a job request is made. This paper challenges that assumption, since not all job requests from a Cloud server can be scheduled on an Edge node. Thus, guaranteeing fairness among the clients (Cloud servers offloading jobs) while accounting for priorities of the jobs becomes a critical task. This paper presents four scheduling techniques, the first is a naive first come first serve strategy and further proposes three strategies, namely a client fair, priority fair, and hybrid that accounts for the fairness of both clients and job priorities. An evaluation on a target platform under three different scenarios, namely equal, random, and Gaussian job distributions is presented. The experimental studies highlight the low overheads and the distribution of scheduled jobs on the Edge node when compared to the naive strategy. The results confirm the superior performance of the hybrid strategy and showcase the feasibility of fair schedulers for Edge computing.


\end{abstract}

\section{Introduction}
\label{sec:introduction}
Edge computing is a relatively recent distributed computing paradigm that leverages resources at the edge of the network for improving the overall performance of an application~\cite{edgecomputing-00, edgecomputing-03, shi-edgecomputing}. Edge resources may include routers or base stations, or dedicated micro clouds located at the network edge. 

Typically, an application may be offloaded from the Cloud to the Edge or from end user-devices to the Edge~\cite{acmresmgmtsurvey-01}. In the former case of offloading, a service hosted in the Cloud is brought closer to user-devices at the Edge for minimising communication latency~\cite{offloadc2e-01, offloadc2e-02} and reducing the volume of data transferred to the Cloud~\cite{fogdata}. The computational resources available on the Edge are hardware limited (small form factor, limited power, processing cores and storage) when compared to large amounts of hardware available in a Cloud data center~\cite{offloadc2e-03}. 
In the latter case of offloading, computational jobs that cannot be processed on battery powered user devices are offloaded to Edge nodes that are powered via main lines and consequently can host more resources with more computational capacity than available on the device~\cite{offloadd2e-01, offloadd2e-03}. 

This paper considers the case when services of an application are offloaded from the Cloud to the Edge; these offloaded services are explicitly referred to as a \textit{`jobs'} in this paper. Edge computing jobs are different from traditional Cloud workloads in that they are not entire applications, rather a subset of the services of an application that may be latency critical/bandwidth intensive. Executing these jobs on the Edge will result in an improvement in the Quality-of-Service (QoS) of the entire application. 
Incoming jobs to the Edge will need to be efficiently scheduled on to the resources. 
Although there is a large body of research that tackles scheduling for different parallel and distributed systems, such as Grids~\cite{gridsched-01}, Clusters~\cite{clustersched-01}, and Clouds~\cite{cloudsched-01}, scheduling on the resource constrained Edge is more challenging~\cite{edgesched-02}. 
%This is because the competition to acquire Edge resources for executing jobs will be significantly more than on the Cloud given the abundance of resources in the Cloud when compared to hardware limited resources at the Edge. 
Due to limited hardware resources available on the Edge, it is not easy to adopt the aforementioned approaches for Edge computing. 

%Ideal scheduling will need to be fair because higher rates of incoming jobs from an application is more likely to be scheduled on the edge, denying access to those jobs (or queuing them) that are not frequent. This motivates the need for fairness in scheduling. There are multiple perspectives to fairness in a computing system and is subjective to the parameters that fairness is based on. 

There are various scheduling strategies that are considered for computing systems. In all cases, 
ensuring fairness is a key prerequisite as many unwanted situations can otherwise occur. Consider for example when higher rates of incoming jobs from a specific application are scheduled on the Edge and deny access (or queue) to the ones that are less frequent. 

A large proportion of existing Edge scheduling research implicitly assumes that all job requests can be successfully scheduled on an Edge node; because resources are available whenever a request is made. However, this paper makes the assumption that not all incoming job requests from a Cloud server can be scheduled on the Edge node. This is because Edge resources are hardware limited. Therefore, there will be significant competition to gain access to these resources. Developing scheduling strategies that ensure fairness among clients, which is the objective of this paper, becomes essential.

Cloud and Edge resources are likely to be owned by different operators or service providers; this translates to Cloud application owners having to pay for using Edge resources~\cite{dyverse-01}. Assuming a cost model in which users pay for Edge resources, priorities will need to be accounted for since users may pay for premium service (higher priority execution) on the Edge~\cite{dyverse-01}. 

Although fair scheduling has been extensively studied in Cloud computing~\cite{wei2010game, wang2015multi}, it needs to be revisited within the context of Edge computing~\cite{fogchallenges}. 
This is because, as already highlighted, Edge resources are hardware limited and therefore it is a very hard task to dynamically auto-scale. 
For example, some edge nodes, such as routers, will already host a traffic-routing service and may only be able to host some offloaded workloads from the Cloud for a small amount of time. 

In this paper, we consider fair scheduling for a large number of relatively short running incoming job requests at the same time to an Edge resource that need to be scheduled by taking priorities of jobs into account without being unfair to any specific type of incoming jobs. 
Given a number of clients that own different jobs, the scheduler proposed in this paper determines which jobs of a client can be scheduled on to an Edge node to ensure fairness among all clients. The research presented in this paper is concerned with the fairness of the scheduling strategy on a single Edge resource. 
Nonetheless, it is noted that the proposed fair scheduling strategies can be easily extended to undertake cooperative and fair load-balancing across multiple Edge resources.

Four scheduling techniques are presented - the first come first serve strategy used as a baseline and three additional techniques are proposed that account for different notions of fairness. They are: (i) a client fair strategy that accounts for the fairness of the client submitting the job, (ii) a priority fair strategy that accounts for the fairness of the priorities of the job, and (iii) a hybrid strategy that accounts for priorities of incoming jobs and the client submitting them. An experimental study is carried out on a target platform to identify the overheads and distribution of jobs. Preliminary investigation confirms the feasibility of the proposed scheduling technique and suggests the superior performance of the hybrid strategy. 
    
This paper is organised as follows. Section~\ref{sec:scheduler} presents the proposed scheduler. Section~\ref{sec:jobmanagement} considers job management by taking fairness of the client and priorities of the job into account. Section~\ref{sec:experimentalstudies} presents experimental results obtained. Section~\ref{sec:relatedwork} discusses the related work. 
Section~\ref{sec:conclusions} concludes this paper by presenting opportunities for future work. 

\section{Fair Scheduling on the Edge}
\label{sec:scheduler}
To effectively employ Edge computing, a research question that needs to be answered is how to schedule different workloads offloaded from multiple Cloud applications such that they can share hardware resources on an Edge node in a fair manner. This section investigates the requirements for fair scheduling in Edge computing and proposes an `Edge Fair Scheduler (EFS)' that can operate at the edge of the network. %The architectural design is presented to understand the fit of EFS within Edge computing. 

\subsection{Architecture}
\begin{figure}[t]
    \centering
    \includegraphics[width=0.37\textwidth]{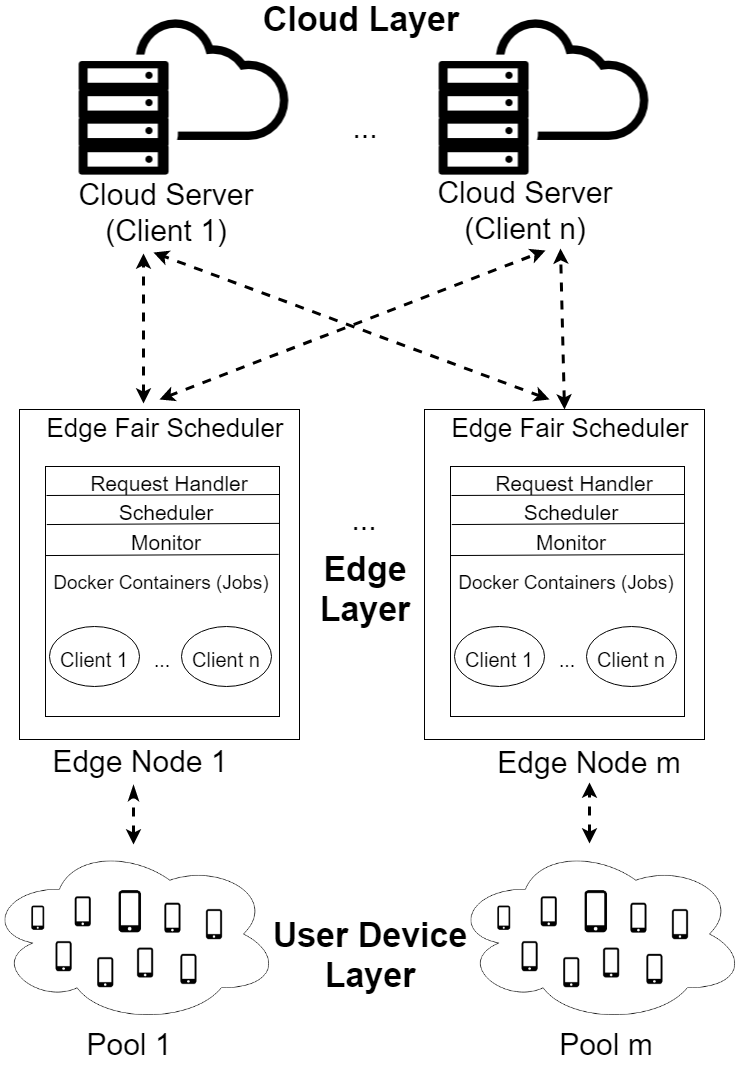}
    \caption{Architecture of EFS in an Edge computing environment}
    \label{fig:arch}
\end{figure}

The proposed EFS is designed to be integrated in a three-layer Edge architecture that consists of the Cloud layer, the Edge layer and the User layer~\cite{arch}. With such an architecture, a part or all of a Cloud application is offloaded from the Cloud layer to the Edge layer when Edge computing services are requested. When the offloaded application is deployed on Edge nodes, the users connect to the Edge nodes instead of the Cloud, which reduces the latency in communication by allowing data to be processed closer to the users. Figure~\ref{fig:arch} shows the EFS architecture. The EFS is required on each Edge node to provide fair scheduling to all workloads that intend to use Edge computing services on this Edge node. 

The Cloud servers will host long-running applications, and in this paper they are considered as clients of Edge computing services. A client script is provided for each Cloud server to: (i) request a new job to be started on an Edge node; (ii) deploy an application onto an Edge node; (iii) terminate an application that is currently running on an Edge node. The applications are hosted in Docker containers for the purpose of isolation. When requesting to use the Edge service on an Edge node, the client can decide to either deploy all or a subset of workloads of the Cloud application using containers.

The user device layer consists of end users devices that communicate with the deployed applications on Edge nodes. The users will be redirected from the Cloud to the Edge node on which the application has been deployed. %As far as the user is concerned this is a seamless transaction that does not affect them. 
The redirection of users from the Cloud to Edge nodes is the responsibility of the clients (i.e. the Cloud server) deploying the applications.

%On the Edge level, the base service represents the primary services running on an Edge node. For example, the base service of a router is traffic routing while the base service of a mobile base station includes digital signal processing. 
The main components of the EFS on an Edge resource comprises: (i) a request handler that listens to new requests from any of the authorised clients, (ii) a scheduler that decides which request will be processed next, (iii) a monitor that detects and terminates idle jobs, and (iv) a number of application servers deployed in Docker containers. The Edge layer will realistically comprise a large number of Edge resources. However, in this prototype of EFS, only a single Edge node is considered. EFS on multiple nodes is not within the scope of this paper.
%The EFS architecture has been carefully designed to provide fast and reliable services. All the components are independent to ensure high performance of the EFS. The failure of one component will not affect or stop the other components of the EFS. 
The EFS components in detail are:

(i) \textit{Request Handler:} this component handles all the incoming client requests to the EFS. It only accepts new connections that have the correct Secure Sockets Layer (SSL) certificates to authorise the client. This allows client identification and tracking of the frequency of client job execution. The handler will then perform the necessary action based on the request type, for instance to start, deploy, or terminate a job. Any new job requests are queued in a database and are then further handled by the Scheduler. Requests to terminate jobs are also queued in the database and then further processed by the Monitor. All invalid requests are ignored and the client will be notified that an invalid request has been made.

(ii) \textit{Scheduler:} performs the scheduling of queued job requests. It monitors the database queue and schedules the next job if the queue is not empty and resources are available. The scheduling strategy is specified by a config file. Based on the selected scheduling strategy a corresponding algorithm is executed to determine which job is the next to be executed. The different scheduling strategies considered in EFS are discussed in the next section. This component will also notify the clients that the requested job is being started and will set up password-less SSH access for the client to the container.

(iii) \textit{Monitor:} this component is responsible for the management of already running job containers on the Edge node. It detects idle containers, which will be queued for termination once detected. The second function is to monitor the termination queue and to terminate any of the queued containers. The containers to be terminated may be the idle, which were detected by the Monitor, or the jobs which have been requested for termination by the client from the Cloud.

(iv) \textit{Job Containers:} this component is the pool of running Docker containers that host the jobs offloaded from the Cloud. These containers provide an isolated and safe execution environment for the applications. Appropriate networking allows for the services in the containers to be accessible by the end users. A client is only allowed to access its corresponding container which is handled by the Scheduler when the container is booting up. Once a container is terminated it is deleted alongside all the data footprint on the node.

\section{Job Management in the Scheduler}
\label{sec:jobmanagement}

\begin{table*}[!t]
%% increase table row spacing, adjust to taste
%\renewcommand{\arraystretch}{1.3}
% if using array.sty, it might be a good idea to tweak the value of
% \extrarowheight as needed to properly center the text within the cells
	\caption{Notation used in the proposed Edge Fair Scheduler}
	\label{tab:notation}
	\centering
	\begin{tabular}{llr}
        \hline
        Symbol & Description & Source\\
        \hline
        service & Flag to indicate if the scheduler is running &\\
        maxjobs & Maximum number of jobs allowed to run at one time on the node &\\
        $Cpu_u$ & Single unit of CPU assigned to each job; each unit can be set by EFS& \\
        $Mem_u$ & Single unit of RAM memory assigned to each job; each unit can be set by EFS& \\
        $C_n$ & Client name & \\
        $C_{ip}$ & Client IP address & \\
        $C_p$ & Client port for communications & Request Handler\\
        $J_p$ & Job priority \\
        $J_{ports}$ & Ports required by the job &\\
        $J_{id}$ & Job ID &\\
        $J_{qmax}$ & Maximum number of jobs allowed in the job queue &\\
        \hline
        $J_q$ & Job queue &\\
        $J_r$ & Number of currently running jobs on the Edge node &\\
        $C_l$ & List of waiting clients in the job queue\\
        $C_f$ & Key-value store containing frequencies per client \\
        $P_l$ & List of priorities of jobs that are waiting in the queue & Scheduler\\
        $P_w$ & Key-value store containing the weights associated with each priority\\
        $P_f$ & Key-value store containing percentages of jobs executed per priority\\
        $P_c$ & Number of jobs executed for a specific priority\\
        Total & Total number of all jobs executed in the past\\
        \hline
        $T_q$ & Termination queue\\
        C & Docker container \\
        $C_r$ & List of running containers \\
        prevstats & Key-value store of CPU usage for each container at start of monitoring period & Monitor\\
        currentstats & Key-value store of CPU usage for each container at end of monitoring period\\
        percentages & Key-value store of the percentages of CPU usage per container \\
        idle & List of idle containers \\
        \hline
    \end{tabular}
\end{table*}

The proposed EFS has three functionalities, namely job request handling, the scheduling of jobs and the monitoring of jobs. Request handling refers to managing incoming requests from authorised clients and handling them. Scheduling jobs is a main task of the EFS and the associated scheduling strategies decide which queuing jobs are next executed. In general, job scheduling on a distributed system is an NP-Complete problem and this assumption holds true for the Edge computing scenario. Monitoring jobs in the containers is to release compute resources from containers that are either idle or requested to be terminated. Table~\ref{tab:notation} presents the notation used in the algorithms presented in this section.

\subsection{Request Handling}

\begin{algorithm}
\SetAlgoLined
\KwData{$service$, $maxjobs$, $J_q$, $J_{qmax}$, $C_n$, $C_{ip}$, $C_p$, $J_p$, $J_{ports}$, $J_{id}$}
$Cpu_u,Mem_u,maxjobs,J_{qmax}$ = read config\;
\While{service == True}{
    accept new connection\;
    get client SSL\;
    \uIf{SSL valid}{
        request = get request\;
        \uIf{request = "New Job"}{
            \uIf{$J_q.length() < J_{qmax}$}{
                $J_{id}$ = queue job($C_n, C_{ip}, C_p, J_p, J_{ports}$)\;
                notify client($J_{id}$)\;
            }
            \Else{
                notify client of no space\;
            }
        }
        \uElseIf{request = "Terminate"}{
            \uIf{$J_{id} in J_q$}{
                remove from queue($J_{id}$)\;
                notify client()\;
            }
            \Else{
                queue for termination($J_{id}$)\;
                notify client()\;
            }
        }
        \Else{
            notify client of invalid request\;
        }
    }
    \Else{
        close connection\;
    }
}
\caption{Request Handling}
\label{alg:req_handler}
\end{algorithm}

Algorithm~\ref{alg:req_handler} demonstrates the request handling process of the EFS on an Edge node. This allows the clients to make job related requests to the Edge node. When the EFS is first started it reads a configuration file which stores the required values such as the maximum number of jobs that can be supported by an Edge node $maxjobs$ (Line~1). Once the configuration is loaded then the request handling service starts running continuously until being terminated by a SIGINT signal. The service listens for any new incoming requests from any of the clients. When a new HTTP socket connection is established (Line~3) the client is then asked for an SSL certificate (Line~4) which should have previously been acquired from the Edge service provider (i.e. Edge node owner) for establishing a secure encrypted connection. The creation and distribution of SSL certificates is the responsibility of the service provider. The provider can either choose to sign the certificates using OpenSSL or a third party SSL certificate signing authority. Since the certificates are distributed to clients individually they assist the identification of which client is connecting. Once a secure connection is established using valid SSL certificates the request handler awaits request from the client (Line~6).
%\cite{openssl} 

There are two valid requests -- to start a new job (Line~7) or to terminate a job (Line~14). If the client requests a new job to be started (Line~7), the service first checks if there is any space left in the queue for anther job (Line~8). If a space is available the job is added to the database, with the job requirements received in the initial request to start this job. Once the job is queued the client is notified of the job ID which can be used in the future to terminate the job (Line~10). If there is no space in the queue, the client is notified of rejection due to the lack of available space (Line~12). If the request is to terminate a job (Line~14), the job will be removed from the queue if it is in the queue (Lines~15-16). Jobs that are already running on the Edge node will instead be queued for termination (Lines~18-19). Then the client is notified that the request has been accepted successfully (Lines~17 and~20).

\subsection{Job Scheduling}
The job scheduler is responsible for firstly identifying the next job to be executed, and secondly provisioning hardware resources of an Edge node for executing the job. The decision of which job is the next job to be scheduled affects the fairness of the Edge resources. Four scheduling strategies are presented in this paper; the first is the First Come First Serve (FCFS), and the three remaining are proposed for EFS, namely the fairness for the clients (referred to as Client Fair), fairness for the jobs (referred to as Priority Fair), and a hybrid strategy.

\subsubsection{First Come First Serve} FCFS is a popularly adopted baseline strategy in Edge research~\cite{fcfs, singh2017rt}. This strategy uses the time sequence of the jobs arrived to select the next job to be executed. When the FCFS strategy is applied, the scheduler considers fairness on the basis of the time of job request submission. It looks up the oldest entry in the job queue, along with the job requirements. This entry in the database is then moved from the queue into the history table. All of the job requirements are then passed to the scheduler for launching an appropriate container for the job. 

The FCFS-based scheduler does not consider where a request is coming from. There may be a case that one user sent 100 jobs to an Edge node slightly before the second user tries to use Edge resources for its first job. With the FCFS-based scheduler, the Edge node would be exhausted by jobs submitted from the first user, leaving the second user in an unfair competition. Therefore, the origin of the job requests (or users of the Edge) need to be considered.

\begin{algorithm}
\SetAlgoLined
\KwData{$J_{id}, C_n, C_{ip}, C_p, J_p, J_{ports}, C_l, C_f$}
$C_l$ = get waiting client list\;
\For{$C_n$ in $C_l$}{
    $C_f[C_n]$ = get client frequency($C_n$)\;
}
$C_n$  = argmin($C_f$)\;
$J_{id},C_n,C_{ip},C_p,J_p,J_{ports}$ = fetch oldest entry in queue for client($C_n$)\;
move job from queue (retrieved in line above) to job history\;
\Return $J_{id},C_n,C_{ip},C_p,J_p,J_{ports}$\;
\caption{Client Fair}
\label{alg:client}
\end{algorithm}

\subsubsection{Client Fair strategy} Different from the FCFS strategy above, this strategy considers fairness for all users of an Edge node. This concept of fairness was initially proposed to ensure that all users of a computing node will have a fair share of the processing resources~\cite{kay1988fair}. In the Edge computing environment targeted in this paper, when the fairness for the client strategy is applied, the scheduler ensures that between all of the clients who request jobs on the Edge node an equal amount of jobs from each client are executed providing sufficient requests are made. This strategy prevents clients who make frequent job requests from being favoured and allows new clients get a chance of using Edge resources.

Algorithm~\ref{alg:client} implements the fairness for the client strategy in the EFS, referred to as the Client Fair algorithm. Initially a list of the currently waiting clients is created (Line~1). For each of the clients in the list, the frequency of job executions is calculated (Line~3). The client whose jobs have been least frequently executed is then selected (Line~5). The oldest entry in the job queue for that client is then selected along with the job requirements (Line~6). The job is then moved from the queue into the job history (Line~7) before the job requirements are returned to the scheduler (Line~8). 

This Client Fair algorithm aims to provide equal frequency of executing jobs for all users on an Edge node. However, it does not distinguish the jobs (i.e. all jobs are assumed as equally important). There could be a case that a client whose jobs are all highly important is not executed on the Edge node because by this algorithm another client whose jobs are not important at all needs to be fairly treated. Therefore, the next scheduling strategy focuses on the importance of jobs by prioritising them. The complexity of the algorithm is $O(1)$, which is a constant assuming that the job queue is up-to-date and queues the job in descending order of job arrival time. 

\subsubsection{Priority Fair strategy} In contrast to the above strategy, this strategy considers fairness on the basis of jobs with different priorities. Algorithm~\ref{alg:priority} demonstrates how fairness is ensured for all jobs on an Edge node, referred to as the Priority Fair algorithm. In particular, three levels of priorities are considered for the jobs to be executed on the Edge node. Firstly the scheduler assigns a weight to each of the priority levels (Line~1). Starting at the highest priority value of 3 to the lowest of 1, the weighting is set as follows: 50\%, 35\%, 15\%. This is necessary because only using the three priority levels may cause a starvation problem for low-priority jobs when there are too many high-priority jobs. By applying weights, jobs with the lowest priority level will have a chance to use Edge resources and meanwhile the probability of executed jobs will still follow the pre-defined three priority levels.

Similarly to the previous Client Fair scheduling strategy, a list of waiting priorities is constructed (Line~3). For each of the priorities in the constructed list, a job frequency percentage is calculated (Lines 5-9). Then the scheduler selects an appropriate priority level, from which a job will be further selected as the next job to be executed. To identify the priority level, the selected job priority is firstly initialised with a value of $-1$ (Line~12). Then starting at the highest priority level, the frequency of jobs having been executed for each priority level is compared against the priority weightings. If a priority level is below the weighting threshold configured, then this priority level is selected (Lines 14-15). Such a searching process for the appropriate priority level is repeated until a valid priority value is obtained. In the case that the waiting priorities list is exhausted, the highest priority level is selected as the next to be executed (Lines~16-17). This is because the priority level is assumed to be purchased by clients and the more times high-priority jobs are executed the more revenue can be generated for the Edge service provider.

\begin{algorithm}
\SetAlgoLined
\KwData{$J_{id}, C_n, C_{ip}, C_p, J_p, J_{ports}, P_c, P_w, P_l, P_f, Total, index=0$}
$P_w$ = set priority weighting\;
$Total$ = get number of all past jobs\;
$P_l$ = get waiting priorities list\;
\For{$J_p$ in $P_l$}{
    $P_c$ = get number of jobs executed for specific priority\;
    \uIf{$P_c > 0$ and $Total > 0$}{
        $P_f[J_p]$ = $P_c/Total$\;
    }
    \Else{
        $P_f[J_p] = 0$\;
    }
}
$J_{p}=-1$\;
\While{$J_{p}==-1$}{
\uIf{$P_f[P_l[index]] < P_w[P_l[index]]$}{
     $J_{p}=P_l[index]$\;
}
\uElseIf{$index = len(P_l)-1$}{
    $J_{p}=P_l[0]$
}
\Else{
    $index++$\;
}}
$J_{id},C_n,C_{ip},C_p,J_p,J_{ports}$ = fetch oldest entry in queue for priority($J_p$)\;
move job from queue to job history\;
\Return $J_{id},C_n,C_{ip},C_p,J_p,J_{ports}$\;
\caption{Priority Fair}
\label{alg:priority}
\end{algorithm}

Once the next priority level of jobs to be executed is selected, the oldest entry in the job queue for that priority level is then selected and its job requirements are retrieved (Line~22). The job request entry is then moved from the job queue into the job history table (Line~23). Following this the job requirements are returned to the scheduler for starting the job (Line~24). The complexity of the algorithm is $O(P)$, where $P$ is the number of priority levels considered in EFS. 

Priority Fair is different from Client Fair as it focuses on the importance of jobs and aims to ensure the relative priorities of jobs while giving a reasonable share of Edge resources to low-priority jobs. However, the priority fair algorithm does not consider where these jobs come from, and thus there is no guarantee of fairness for different users of an Edge node. A hybrid fair scheduling strategy is developed to consider the fairness of both the clients and the jobs, 

\subsubsection{Hybrid strategy} To provide fairness for both the clients and the priority of the jobs at the same time, this Hybrid Fairness scheduling strategy combines the design of the Client Fair and the Priority Fair strategies, as illustrated in Algorithm~\ref{alg:hybrid}.  When this strategy is applied, the scheduler firstly identifies the appropriate priority level of jobs to be executed next (Line~1). Then instead of directly choosing the oldest entry in the job queue for that priority level, the Hybrid Fairness scheduler calculates the frequency of executed jobs for all clients. Only jobs submitted by the client having the least number of jobs being executed is then further considered (Line~2). Finally the oldest entry in the job queue for the selected priority level and the selected client is scheduled to be executed next (Lines~3-4). By doing so the probability of executing high-priority jobs is kept and the clients are satisfied because no users of an Edge node will receive a larger share of the resources than others. The complexity of the algorithm is the sum of the complexities of the priority fair and client fair algorithms presented above. 

\begin{algorithm}
\SetAlgoLined
\KwData{$J_{id}, C_n, C_{ip},C_p, J_p, J_{ports}, P_c, P_w, P_l, P_f, Total, \linebreak C_l, C_f$}
run Lines~1-21 of the Priority Fair algorithm\;
run Lines~1-5 of The Client Fair algorithm\;
$J_{id},C_n,C_{ip},C_p,J_p,J_{ports}$ = fetch oldest entry in queue($J_p$, $C_n$)\;
move job from queue to job history\;
\Return $J_{id},C_n,C_{ip},C_p,J_p,J_{ports}$\;
\caption{Hybrid}
\label{alg:hybrid}
\end{algorithm}

After the next job to be executed is identified, the scheduler provisions Edge resources for the job and continuously polls the database to check if there are any queuing jobs. A new job will be executed if the maximum number of jobs that can be supported on the Edge node is not exceeded and there are resources available on the Edge node. The selection of this new job is done by the EFS using one of the four scheduling strategies presented above. Each strategy considers fairness in a different way to demonstrate that fairness is subjective. 

Once the next job has been selected a container is started and all the requested ports are forwarded. The container automatically binds the ports to all available interfaces on the Edge node. An SSL session is established with the client. Then the client is notified of the container being started along with a key-value store of the port mappings. In return, the client sends the EFS a public SSH key that is appended to the authorised hosts file within the container for secure access. Following this set-up the client installs any dependencies, copies necessary files and starts the application in the container.

This section presents job scheduling in the EFS. 
The scheduling strategies presented in this paper are executed on the Edge nodes and they have been carefully designed to not exhaust resources on the Edge node. 
The jobs to be executed on the Edge are expected to be short-running jobs due to the dynamic nature of the Edge computing environment~\cite{offloadc2e-01}. Therefore, the EFS Monitor is presented in the next section to control the life cycle of the running jobs on an Edge node.

\subsection{Monitoring and Terminating}
The Monitor has two functions. Firstly, it monitors all running containers on an Edge node and identifies inactive ones for termination in a later stage, as shown in  Algorithm~\ref{alg:stop_idle}. To not overload the Edge node, the monitoring process is performed every two minutes (Line~20). Each time the monitoring process is carried out, CPU statistics of all the containers running for at least one minute is gathered (Line~4). The one minute period allows clients to set up and deploy Edge applications in the container. There is a 10-second interval between the initial and updated CPU measurements for all containers that have been running for at least one minute (Line~9). The CPU utilisation (in \%) is obtained for all containers that have valid CPU statistics in the key-value stores previously and currently (Line~11). Containers with a utilisation of less than 10\% are appended to a list of idle containers (Lines 12-14). Containers in the idle containers list are queued for termination (Lines 16-19).

\begin{algorithm}
\SetAlgoLined
\KwData{$service, C, C_r, prev stats, current stats, \linebreak percentages, idle$}
\While{service == True}{
    \For{C in $C_r$}{
        \uIf{C.uptime $>=$ 1 minute }{
            prevstats[C] = get CPU stats(C)\;
        }
    }
    wait 10s\;
    \For{C in $C_r$}{
        \uIf{C.uptime $>=$ 1 minute }{
            currentstats[C] = get CPU stats(C)\;
        }
    }
    percentages = compare stats(prevstats[], currentstats[])\;
    \For{C in percentages[]}{
        \uIf{$percentages[C] < 10.0$}{
            idle.append(C)\;
        }
    }
    \uIf{$idle.len() > 0$}{
        \For{C in idle[]}{
            queue for termination(C)\;
        }
    }
    sleep 2 minutes\;
}
\caption{Stop Idle}
\label{alg:stop_idle}
\end{algorithm}

Secondly, the Monitor stops containers in the termination queue, either because of the Stop Idle process described above or requested by the client directly. %The Monitor continuously checks if any jobs have been lined up for termination. 
When terminating a job, the container is stopped with a default timeout of 10~seconds before forced termination by the SIGKILL command and removed. The termination request is then removed from the termination queue and the client is notified of the successful termination. The complexity of Algorithm~\ref{alg:stop_idle} is $O(p)$ where $p$ is the number of active containers. 

\section{Experimental Studies}
\label{sec:experimentalstudies}
The experimental set up, the scenarios used for evaluating the fairness strategies, and the results obtained from the experimental studies are presented in this section. 

%\subsection{Experimental Setup}
\textbf{\textit{Experimental Setup}} - 
The experimental hardware platform and the evaluation scenarios are considered. 

\subsubsection{Experimental Platform}: The experimental setup for the proposed fair scheduler includes (i) a cloud virtual machine (VM) that uses an eight core Intel Xeon CPU at 2.40 Ghz processor and 16 GB RAM running Ubuntu 16 LTS, (ii) an edge resource that is an Odroid~C2
%\footnote{\url{https://magazine.odroid.com/wp-content/uploads/odroid-c2-user-manual.pdf}} 
 single board computer 
%(as shown in Figure~\ref{fig:odroid}) 
 that has ARM Cortex-A53 (ARMv8) 1.5 Ghz quad core CPUs, 2 GB DDR3 SDRAM, running Ubuntu 16.04. A 16 GB microSD card is used to install the OS and the software is developed in Python 3.5. SQLite 3.11 is used for the database required for the job entries. Docker 3.6 is used for the containers.

%\begin{figure}
%\centering
%\includegraphics[width=0.4\textwidth]{images/odroid.jpg}
%\caption{The Odroid C2 single board computer used as the Edge node in the research presented in this paper}
%\label{fig:odroid}
%\end{figure}

\subsubsection{Workload distribution scenarios}
The experiments assume six clients that can make job requests.
The jobs submitted to the scheduler may be assigned the following three priorities: (i) Priority level 1 (PL1), which is the lowest priority that can be assigned to a job. Jobs with this priority occupy 15\% of the hardware resources, (ii) Priority level 2 (PL2), which is a regular priority and jobs with these priority occupy 35\% of the hardware resource on the edge, and finally (iii) Priority level 3 (PL3), which is the highest priority and jobs with this priority occupy 50\% of the hardware resource. 

Three scenarios using the above priorities are considered for evaluating the scheduling strategies:

(i) \textit{Scenario 1 - Equal job distribution}: each client requests 50 jobs comprising 17 PL1 and PL2 requests and 16 PL3 in a testing window of one hour.   

(ii) \textit{Scenario 2 - Random job distribution}: each client requests up to a maximum of 50 jobs generated at random time intervals. There is no prior knowledge of the rate of arrival of jobs. The results will be evaluated on the amount of jobs executed per client and for each priority as indicated above.  

(iii) \textit{Scenario 3 - Gaussian job distribution}: assuming there are six clients, the clients make jobs as per a gaussian distribution as shown in Table~\ref{tab:gaussian}. The total requests are made within a one hour window. Clients A and B start making requests in the second half of a one hour window. Clients C and D make requests continuously during the test window. Clients E and F make their last requests at the beginning of the test window. Priorities are randomly assigned to the jobs. 

\begin{table}[]
\centering
\caption{No. of job requests per client based on a gaussian distribution}
\begin{tabular}{|l|c|c|c|c|c|c|}
    \hline
    \textbf{Client}          & A  & B  & C  & D  & E  & F  \\ \hline
    \textbf{No. of job requests} & 15 & 45 & 90 & 90 & 45 & 15 \\ \hline
\end{tabular}
\label{tab:gaussian}
\end{table}

\subsubsection{Assumptions}
In this paper, we make the following four assumptions for the experimental studies:
%\begin{itemize}[leftmargin=0.3cm]
    %\item
(i) The scheduling strategies operate for a single Edge resource. The execution of the scheduling strategies on multiple Edge resources will be considered in the future. 
    %\item 
(ii) Each job requires the same amount of CPU and memory resources on an Edge node. This is assumed since we use synthetic workloads to evaluate the feasibility of EFS. To change the resource allocation for individual incoming job requests, the corresponding data for each job will need to be included in a configuration file of EFS.
    %\item 
(iii) Each individual job does not execute for more than 10 minutes on the Edge node. This is assumed because workloads offloaded from the Cloud to the Edge are not anticipated to be long running~\cite{tortonesi2018taming}.  
    %\item 
(iv) EFS has three priority levels for jobs as considered above. The priorities in the experiments presented aim to highlight how priorities are taken into account. Additional priorities can be defined in EFS. 
%\end{itemize}

%\subsection{Results}
\textbf{\textit{Results}} - 
The results obtained from the experimental studies are organised as follows. Firstly, the overheads of the scheduling strategies are considered. Next, the results for the different scenarios considered above, namely for equal, random and Gaussian job distributions, are presented. The results highlight the number of job requests of each client serviced and the percentage of jobs executed based on priorities by the scheduler. 

\subsubsection{Overheads}
\begin{figure}[t]
\centering
    \includegraphics[width=0.49\textwidth]{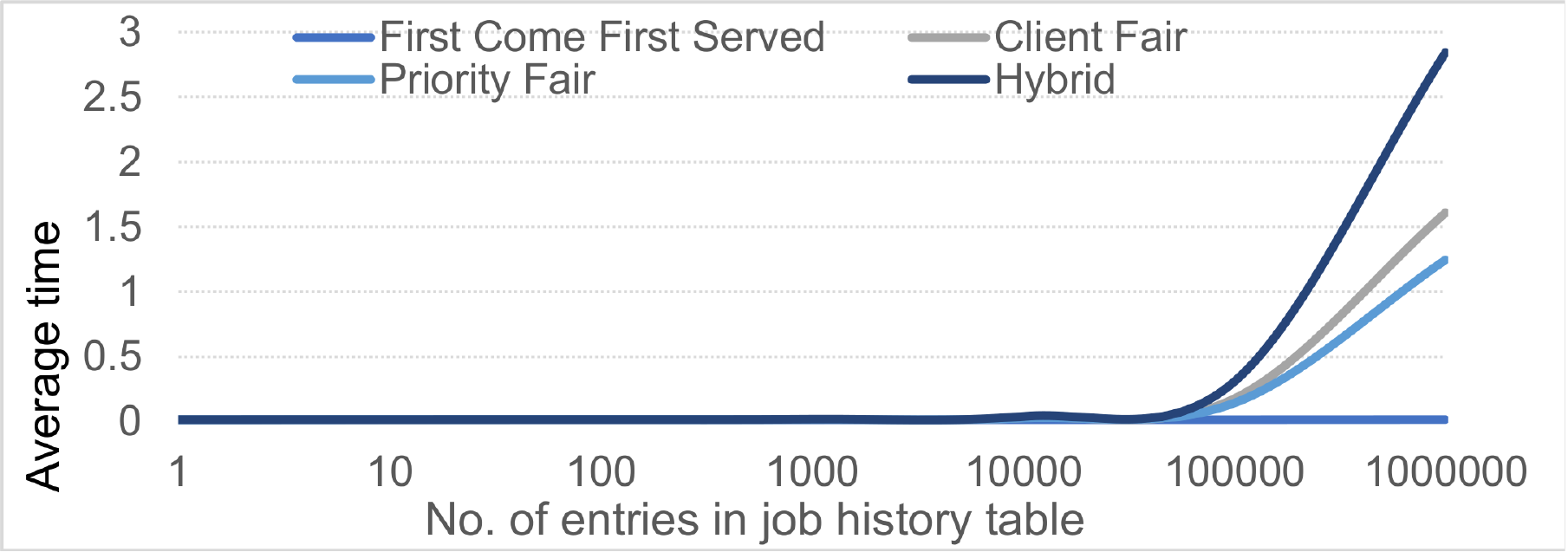}
    \caption{Overheads observed for the fair scheduling strategies with respect to the number of job entries}
    \label{fig:overheads}
\end{figure}

The overheads associated with each scheduling strategy are firstly considered. A key contributing factor to overheads are the database operations to manage the job history, which increases as more entries are added to the job history. The overheads obtained for up to a million entries in the job history are shown in Figure~\ref{fig:overheads}. 

It is immediately observed that the hybrid scheduling strategy has the largest overhead of 2.8 seconds for a job history of a million entries. This is because the number of database look-ups in the job history is twice when compared to the other strategies. The Client Fair (CF) and Priority Fair (PF) strategies both have an overhead of approximately 1.5 seconds (they require only a single look-up). The First Come First Serve (FCFS) strategy has constant overhead of 0.015 seconds as it retrieves the oldest entry in the job queue.

If it is assumed that the time between scheduling jobs is in the order of minutes, then the current overheads would not significantly impact scheduling. The results above are obtained on a single board computer, which in the experimental set up is an Edge node. The overheads can be reduced further if there are additional resources available for scheduling on the edge. 

\subsubsection{Scenario 1 - Equal job distribution}
\begin{figure*}[htp]
\begin{center}
	\subfloat[No. of jobs executed by each client]
	{\label{fig:scenario1a}
	\includegraphics[width=0.475\textwidth]
	{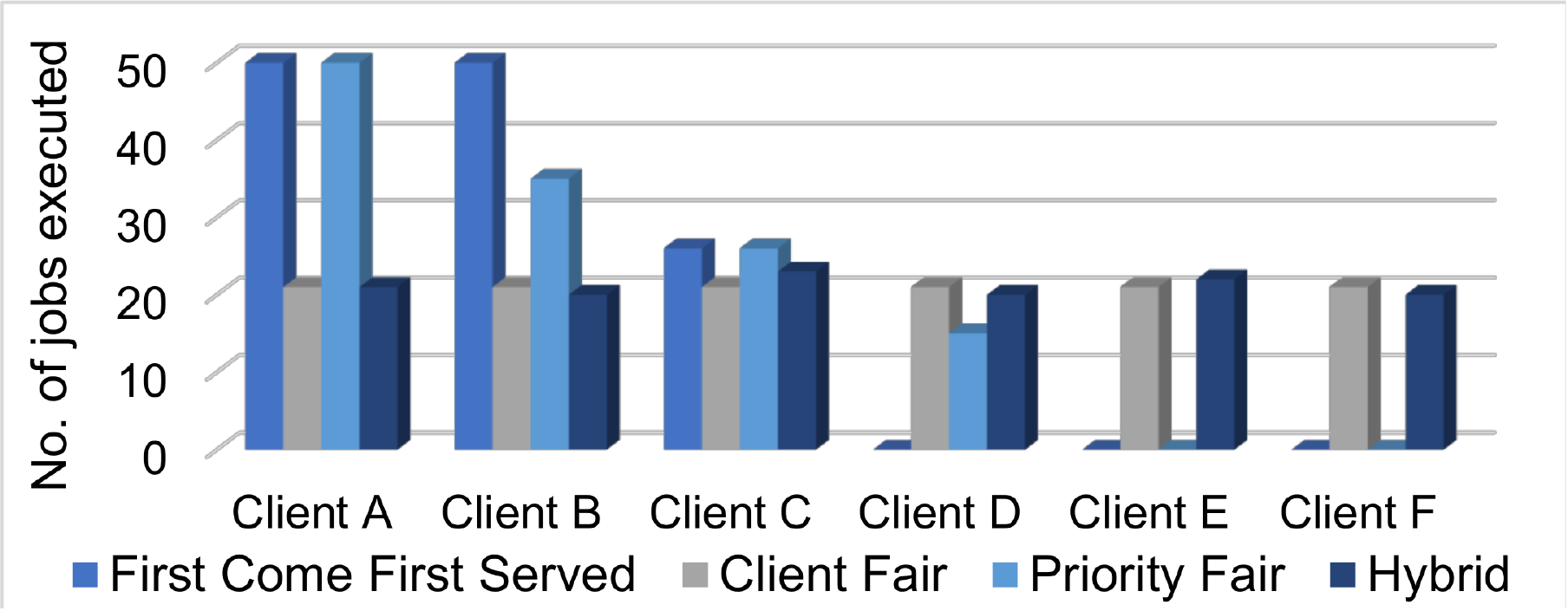}}
\hfill
	\subfloat[Percentage of jobs executed based on priority]
	{\label{fig:scenario1b}
	\includegraphics[width=0.475\textwidth]
	{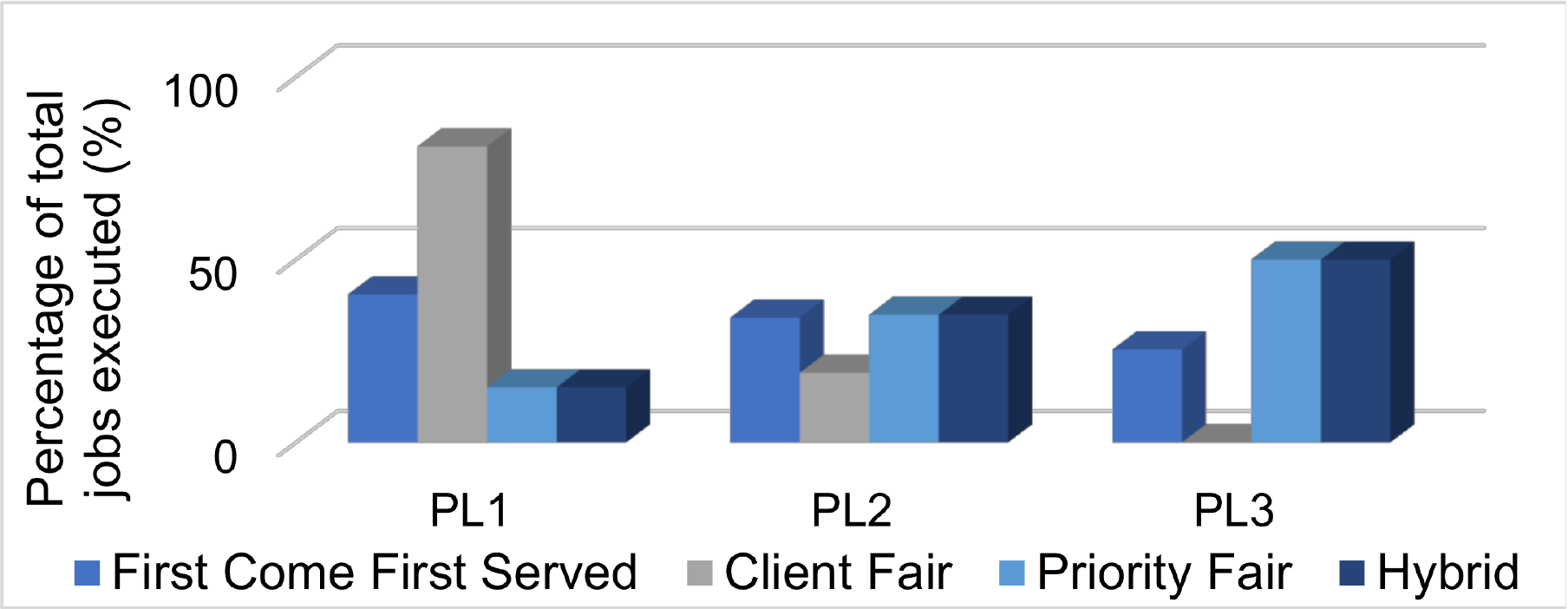}}
\end{center}
\caption{Jobs executed on each client and based on priority in Scenario 1}
\label{fig:scenario1results}
\end{figure*}

Figure~\ref{fig:scenario1a} and Figure~\ref{fig:scenario1b} show the number of jobs and the percentage of jobs executed based on priorities when each client makes an equal number of job requests. The FCFS strategy does not schedule jobs in a fair manner as it selects the first job request and executes it. If one client made a large number of requests during the testing window, then a large proportion of jobs from this client would be executed. No job requests from Clients D, E, and F were satisfied. Although PL3 is the highest priority more jobs from PL1 are executed than PL3.

In relation to the client fair strategy it is observed at the end of the testing window that an equal distribution was achieved among all clients. 
In the client fair strategy, no jobs with priority PL3 were executed. 
If we consider the priority fair strategy, then a larger proportion of jobs from Client A were executed compared to the others; there is a steady decrease in the number of jobs executed and no jobs from Client E or Client F were executed. However, jobs with higher priorities were executed; 50\% of jobs with PL3 were executed, when compared to the 15\% and 35\% of jobs with PL1 and PL2 respectively. Both the client fair and priority fair strategy have either not executed jobs of clients or have completely executed jobs with certain priorities. This is not ideal for a scheduler.

Considering the hybrid strategy for which it is observed that jobs from all clients were executed and jobs with a higher priority (PL3) have been executed in larger proportions when compared to PL1 and PL2. This strategy achieves the combined benefit of the client fair strategy (all clients executed their jobs) and of the priority fair strategy (larger number of jobs with higher priorities are executed). 

\begin{figure*}[ht]
\begin{center}
	\subfloat[No. of jobs executed by each client]
	{\label{fig:scenario2a}
	\includegraphics[width=0.475\textwidth]
	{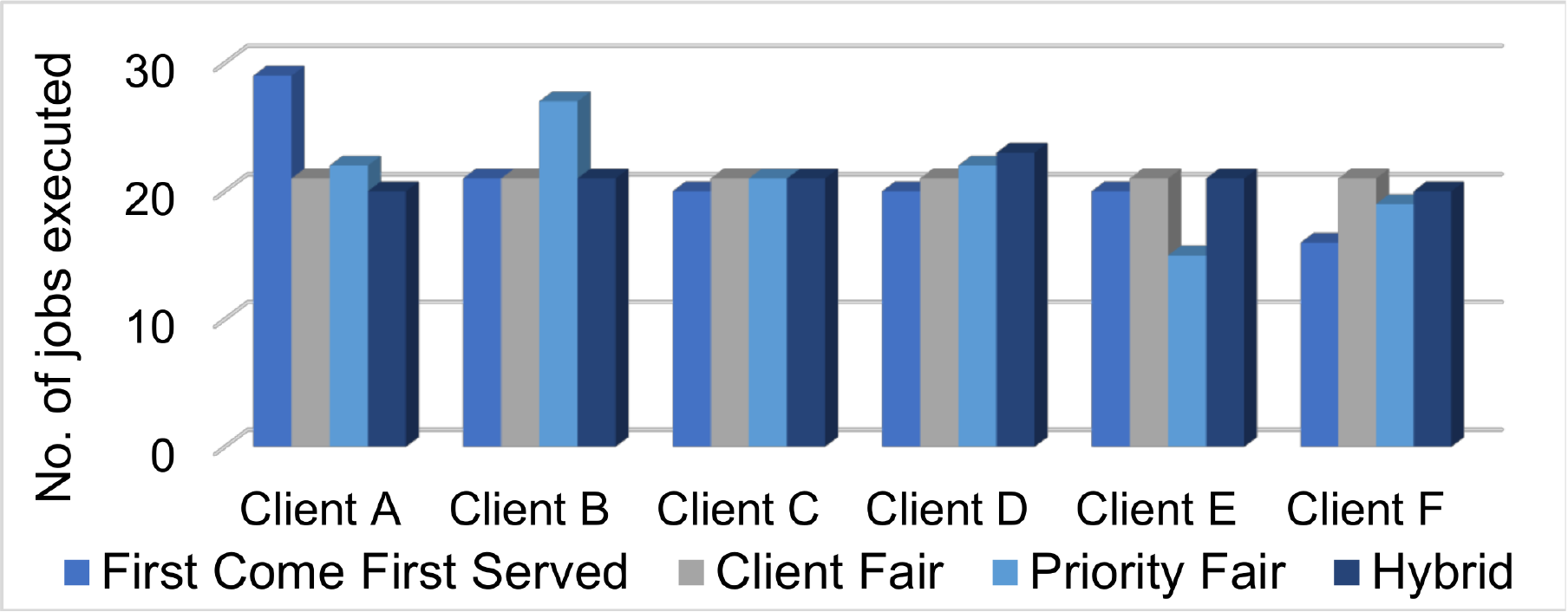}}
\hfill
	\subfloat[Percentage of jobs executed based on priority]
	{\label{fig:scenario2b}
	\includegraphics[width=0.475\textwidth]
	{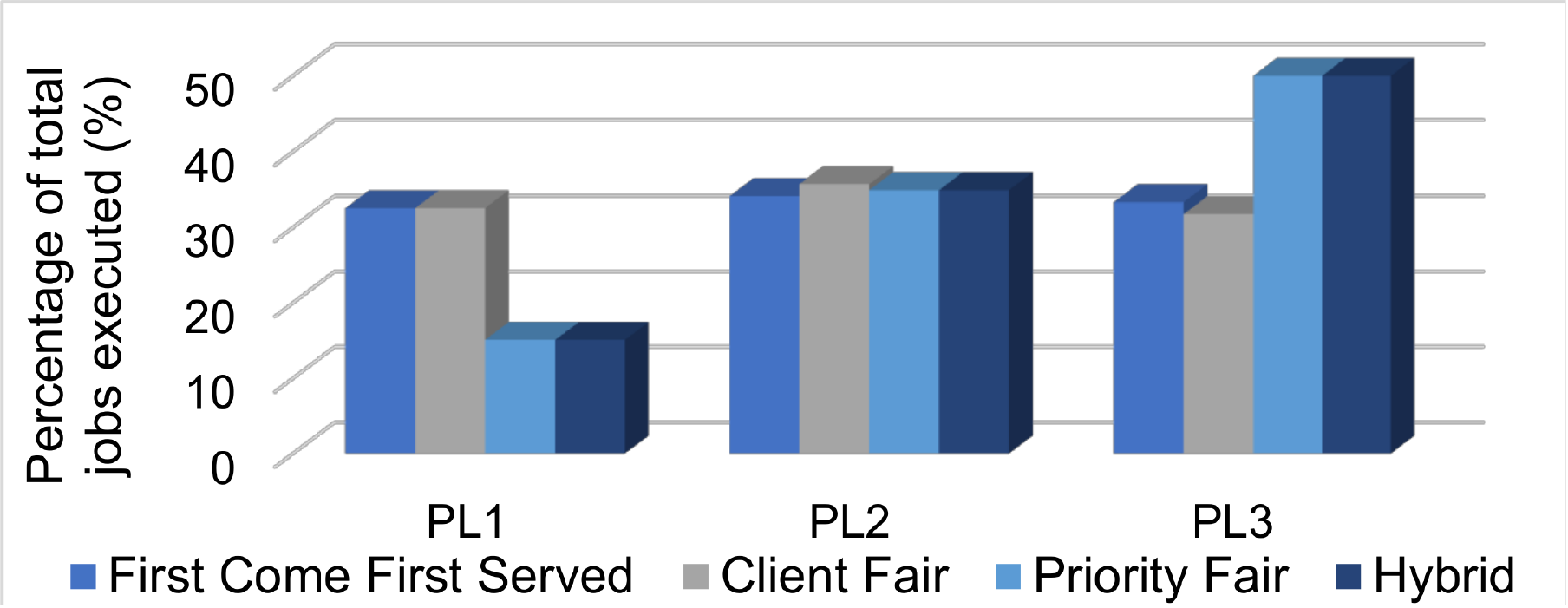}}
\end{center}
\caption{Jobs executed on each client and based on priority in Scenario 2}
\label{fig:scenario2results}
\end{figure*}

\begin{figure*}[ht]
\begin{center}
	\subfloat[No. of jobs executed by each client]
	{\label{fig:scenario3a}
	\includegraphics[width=0.475\textwidth]
	{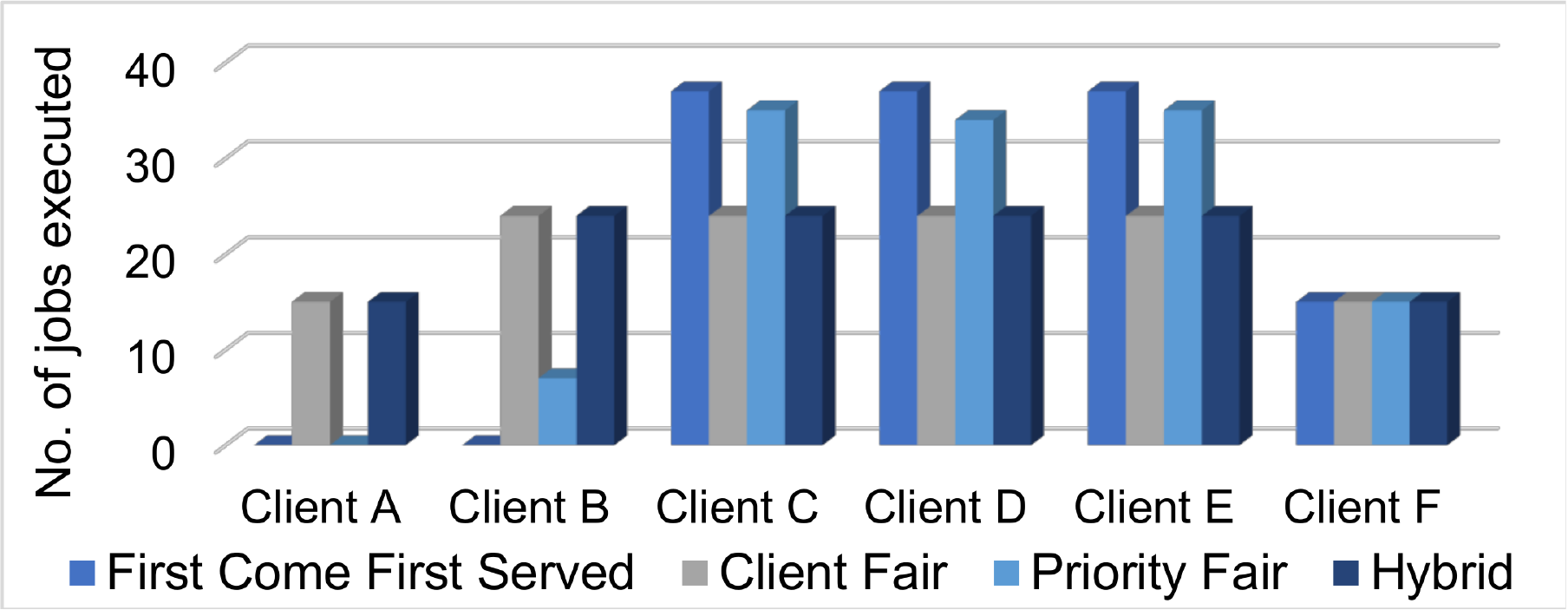}}
\hfill
	\subfloat[Percentage of jobs executed based on priority]
	{\label{fig:scenario3b}
	\includegraphics[width=0.475\textwidth]
	{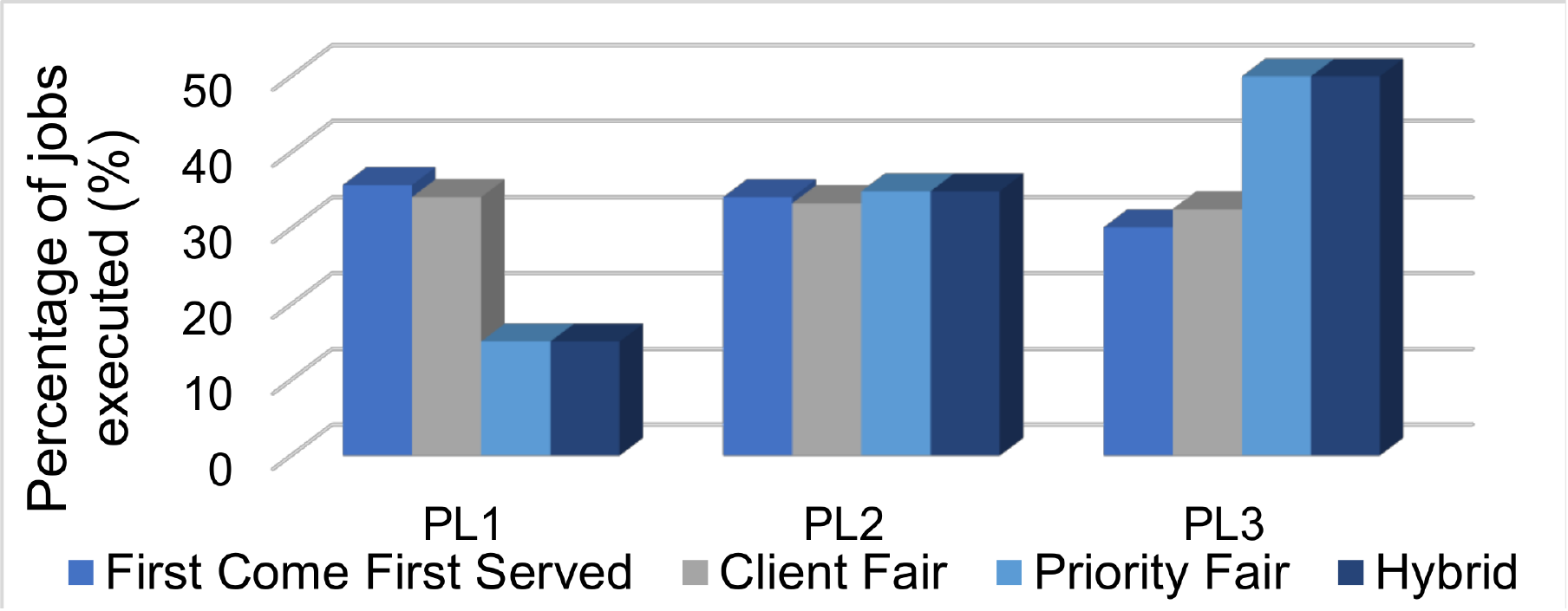}}
\end{center}
\caption{Jobs executed on each client and based on priority in Scenario 3}
\label{fig:scenario3results}
\end{figure*}

\subsubsection{Scenario 2 - Random job distribution}
Figure~\ref{fig:scenario2a} and Figure~\ref{fig:scenario2b} show the number of jobs executed and the percentage of jobs executed based on priorities when each client makes a random number of job requests. It is observed that there is more similarity in the outcome for different strategies. This may be because the random pattern generated for each client over time is the same. In the FCFS strategy, Client A has made more requests than other clients and hence a larger number of jobs is executed. Client F has fewer jobs executed. An equal percentage of jobs with different priorities have been executed. This would naturally be unfair given that higher priorities have not been executed more frequently. 

Considering the client fair strategy, it is observed that the number of jobs executed are nearly the same for all clients. The percentage of jobs executed for different priorities is nearly the same as the FCFS strategy. Although the FCFS and client fair strategies execute nearly equal number of jobs from all clients, they do not account for jobs of different priorities. The priority fair strategy on the other hand executes more jobs with higher priorities, but at the same time has a larger variation in terms of the number of jobs executed for different clients. 

The hybrid strategy (similar to Scenario 1), shows that an equal number of jobs are executed for different clients with a smaller variation, and also executes more number of jobs with a higher priority. When compared to other approaches, the hybrid strategy is more fair for all clients and priorities. 

\subsubsection{Scenario 3 - Gaussian job distribution}
The requests in Scenario 3 are distributed less evenly both in terms of the job requests and the time when the requests arrive. Figure~\ref{fig:scenario3a} and Figure~\ref{fig:scenario3b} shows the number of job requests scheduled and the percentage of jobs executed for different priorities. Only a small proportion of jobs from Client A and Client B are scheduled in the FCFS strategy. Similarly, FCFS is less capable of discriminating between different priorities; eventually fewer jobs with higher priorities are executed. 

The client fair strategy demonstrates more fairness across jobs executed for all clients. However, with regard to the percentage of jobs with different priorities, it is similar to FCFS; fewer jobs with higher priorities are executed. The priority fair strategy in contrast to client fair executes very few jobs of Client A and Client B while executing a higher proportion of Client C, Client D and Client E. A higher proportion of jobs with higher priorities are executed. 

The hybrid strategy achieves a better balance, similar to observations for previous scenarios. However, even with the best approach considered, more than 50\% of the jobs of Client C and Client D are not executed (the Odroid-C2 cannot accommodate all the jobs). Further optimisations will be required to maximise the throughput of the scheduling strategies, which is not considered in this paper.

\subsubsection{Summary}
It is observed that EFS is suitable for hardware limited Edge resources since there are negligible overheads for up to a 100,000 past job entries. For a million job entries and beyond the overhead is 2.8 seconds for the hybrid strategy. This could be ignored if the time between scheduling jobs is in the order of minutes. For any job distribution, the hybrid strategy has superior performance by combining the benefits of both the client fair and priority fair strategies.

\section{Related Work}
\label{sec:relatedwork}
Scheduling is commonly explored in distributed systems to find a mapping between job requests and resources. Numerous strategies are proposed for clusters~\cite{clustersched-01}, clouds~\cite{cloudsched-01}, and more recently for cloud-edge (Fog) systems~\cite{edgesched-02}. The aims of scheduling are load balancing~\cite{edgesched-03}, maximising resource utilisation~\cite{edgesched-06} and energy efficiency~\cite{edgesched-07}, optimising execution costs~\cite{edgesched-04}, and maximising performance~\cite{edgesched-05}. 

The need for scheduling arises when services need to be successfully offloaded. In edge computing research, services can be offloaded either from the user devices to the edge or from the cloud to the edge~\cite{acmresmgmtsurvey-01}. In the case of offloading from user devices to edge resources, there is recent research that develops mathematical models for reducing the job response time, which is defined as the time between releasing a job to an edge node and the arrival of results on the user-device~\cite{edgesched-01}. 

The scheduling strategies required for offloading from the user devices to the edge is vastly different from offloading services from the cloud to the edge. The assumptions are different in both cases. For example, it is unlikely that the same user device would need to offload multiple jobs at the same time. In the research presented in this paper, the focus is on the scheduling strategy required when offloading occurs from the cloud to the edge. The same cloud server (client) can offload multiple short running services (jobs) on to the client. 

In addition, it is anticipated that edge resources will be resource limited when compared to clouds. Therefore, if there is a large influx of job requests from clients (cloud servers for offloading), then given the limitation of resources there will be competition for resource allocation, which requires that there is fairness in the scheduling strategies. Existing edge scheduling research does not consider this, but is the focus of this paper. 

A score-based edge scheduling algorithm for video streaming is proposed to determine the best mapping between services and resources~\cite{edgesched-02}. Cloud solutions and content delivery networks are considered, but assumes that edge resources are always available and do not consider fairness. 

Sample-based scheduling is a technique for load balancing. For this, multiple edge resources are probed to find the existing workload before scheduling another application~\cite{edgesched-03}. Energy efficiency is a parameter considered for scheduling. This is explored in the context of edge micro-clouds~\cite{edgesched-07}.

Application priorities are used for edge scheduling by assuming all job requests can be scheduled~\cite{edgesched-04}. Priorities are used for optimising against response time and overall running costs. However, fairness of the jobs is not considered. 

Evolutionary approaches to scheduling tasks on the edge is proposed, specifically bag of task applications~\cite{edgesched-05}. The parameter optimised is application performance and overall costs. Large amounts of computational resources are required to run the evolutionary approaches and are hence unsuitable to be run on the edge. The scheduler of an edge node will need to be located outside the node or will need to have significant amount of resources. The scheduling strategies presented in this paper operate from within resource constrained resources and ensures fairness to clients submitting jobs to the edge.

Containers are explored for scheduling on the edge given that they are lean deployment mechanisms~\cite{edgesched-06}. Although resource utilisation is maximised, fairness is not considered. 

Existing research considered in this section implicitly assumes that all job requests made for execution on an edge resource can be successfully scheduled; because resources are available whenever a request is made. This paper makes the assumption that not all incoming job requests from a cloud server can be scheduled on the edge node. This is because edge resources are resource constrained and therefore there will be significant competition to gain access to these edge resources. Existing research used for scheduling on the edge does not consider fairness among the clients or the priorities of the incoming jobs. Given a number of clients that own different jobs, the scheduler proposed in this paper determines which jobs of a client can be scheduled on to an edge node to ensure fairness among all clients while considering their fairness. The research presented in this paper is concerned with the fairness of the scheduling strategy on a single edge resource. 

\section{Conclusions}
\label{sec:conclusions}
In this paper, the problem of scheduling jobs of different clients from the cloud on a single edge server is presented. To the best of our knowledge, fair scheduling in edge computing is not explored properly although it has been investigated for traditional computing systems. It is essential that fair scheduling is revisited for edge computing to meet the paradigm specific challenges. This paper presented four scheduling strategies, including the first come first serve, client fair, priority fair, and hybrid. It was demonstrated that each strategy has pros and cons, while the hybrid approach appears to be more suitable as it combines fairness both in terms of the clients and the (given) priorities of jobs to be offloaded. The strategies were tested in three different workload profiles having equal, random and Gaussian job distributions in a test bed comprising a single board computer as an edge resource. 

The current research is a first attempt towards fairness on Edge computing systems. The performance of EFS on a single Edge node is evaluated. Scheduling across multiple Edge nodes can enable load balancing (not considered in this paper). In addition, synthetic workloads with similar execution times under different distribution scenarios are considered. Workloads of varying execution time will need to be considered. 

In the future, it will be formally shown that: (i) the client fair strategy asymptotically serves equally the clients, (ii) the asymptotic serving percentages of the priority fair approach coincide with the preassigned priorities. Further, we aim to explore the notion of fairness for the case where priorities are assigned to clients as well as jobs, and address the challenge of fairness in scheduling having as an additional degree of freedom  the choice to offload to multiple edge servers.

\bibliographystyle{IEEEtran} 
\bibliography{references}

\end{document}